# Poster: Draw a line on your PDA to authenticate

Xiyang Liu, Zhongjie Ren, Xiuling Chang, Haichang Gao
Software Engineering Institute
Xidian University
Xi'an, Shaanxi 710071, P.R.China
{xyliu,hchgao}@xidian.edu.cn

Uwe Aickelin
School of Computer Science
The University of Nottingham
Nottingham, NG8 1BB, U.K.
uxa@cs.nott.ac.uk

## 1. INTRODUCTION

The trend toward a highly mobile workforce and the ubiquity of graphical interfaces (such as the stylus and touch-screen) has enabled the emergence of graphical authentications in Personal Digital Assistants (PDAs) [1]. However, most of the current graphical password schemes are vulnerable to shoulder-surfing [2,3], a known risk where an attacker can capture a password by direct observation or by recording the authentication session. Several approaches have been developed to deal with this problem, but they have significant usability drawbacks, usually in the time and effort to log in, making them less suitable for authentication [4, 8]. For example, it is time-consuming for users to log in CHC [4] and there are complex text memory requirements in scheme proposed by Hong [5]. With respect to the scheme proposed by Weinshall [6], not only is it intricate to log in, but also the main claim of resisting shoulder-surfing is proven false [7]. In this paper, we introduce a new graphical password scheme which provides a good resistance to shoulder-surfing and preserves a desirable usability.

The proposed shoulder-surfing resistant scheme CDS (Come from DAS and Story) inspired by two representative graphical password schemes: DAS [2] and Story [3]. DAS allows users to draw a free-form picture on $N \times N$ grid to produce a password [2] and Story requires users to select a sequence of images to make a story [3]. CDS adopts a similar drawing input method in DAS and inherits the association mnemonics in Story for sequence retrieval. It requires users to draw a curve across their password images orderly rather than click directly on them. The drawing passes through both password images and decoys, which used to confuse peepers. To avoid revealing the first and last pass-images, the drawing must begin and end with given random images. Other complementary measures, such as displaying the degraded images, limiting the length of drawing trace, and erasing the drawing trace are also deployed to strengthen the security.

A preliminary user study was carried to explore the usability of CDS. The result was encouraging that users were able to enter their passwords in 13.7 seconds on average, which is a good result comparing to other recognition based schemes. In comparison to Story, our scheme had a similar memorability, probably due to the same association mnemonics.

## 2. THE PROPOSED SCHEME

The proposed scheme keeps most of the advantage of Story and achieves stronger security. Like Story, our scheme is based on recognition, and suggests users to create a story for sequence retrieval. Instead of direct input, it depends on users drawing a curve across their password images (pass-images) in order.

The scheme uses a set of images gathered from http://images.google.com. The images covering objects, places and people are carefully selected to motivate users' imagination and resized to have identical aspect ratios. To create a password, a user orderly chooses several images from the set as his/her pass-images. The user can remember the connection between the pass-images by mentally constructing a story. For confirmation, the user should draw a curve to cross his/her pass-images in right order without lifting the stylus from the drawing surface.

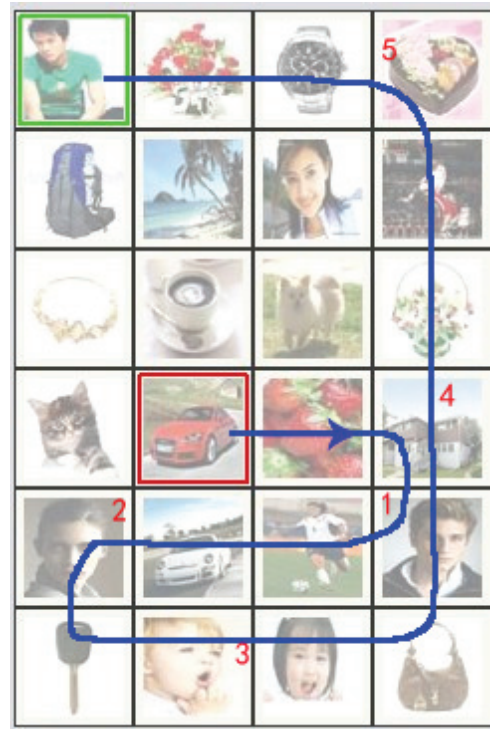

**Figure 1. A possible drawing trace with length 19.**

During the authentication phase, a degraded version of images is randomly displayed on the screen. These degraded images are formed by increasing brightness and reducing contrast of the original images. User must recognize his/her degraded pass-images and draw a curve to orderly cross them. The curve thus passes through both pass-images and other random images used to confuse the attacker. To avoid revealing pass-images (considering a user usually starts with the first pass-image and ends with the last), the drawing must begin with a given random image (head image) determined by the system and end likewise with another (tail image). Moreover, the majority of the drawing trace will be cleared away as the stylus being sliding and only the tail part will be kept to show the current location of the stylus.

Figure 1 shows the interface of the scheme, which uses a template of 24 identically sized images, grouped into a *4×6* matrix. A possible drawing to log in the scheme is shown in figure where the head is labeled by red rectangle and the tail green.

The trace length measured in the number of images crossed is limited within a tolerance in view of the random guess and brute force attacks. The tolerance is adjustable and roughly defined by the maximum length of drawing trace $L$ in our prototype. In default, the maximum length of drawing trace $L$ is defined in terms of the number of images in width $w$ and in length $l$ as well as the number of pass-images $n$ by: $L = (w+l) \cdot (n+1)$. Therefore, *L=(4+6)×(5+1)=60* in Figure 1.

## 3. PRELIMINARY USER STUDY

To discuss the usability of CDS, twenty university students are invited to the study. We conducted a between-subjects design to benchmark the usability against that of Story [3]. Half of the subjects were assigned to the CDS group and half to the Story group. For both group, each participant was required to select five images as his/her pass-images to the purpose of comparing the between group using time.

In CDS group, seven of the participants accomplished the criterion in ten attempts with no errors. The other three participants made a total of four incorrect logins. The mean success login rate is 96.5% (Standard Deviation, StdDev=5.98). In Story group, eight participants made no mistake in ten login attempts and each of the other made one incorrect login. The mean success rate is 98.2% (StdDev=3.83), higher than that of CDS group. However, it is useful to note that in practical terms, this was only a difference of two incorrect responses and should be treated with caution.

The mean time for correct password inputs was also analyzed. As shown in Table 1, the participants in CDS group took more time to log in than their counterparts in Story group on average (13.7s vs. 9.2s). A t-test yields a result of t=3.91, p<0.01(two tails), indicating that there was statistically significant difference between two conditions.

**Table 1. Login time (seconds) for all correct inputs**

| Group | Avg. | t-test | S.d. | Max | Min |
|---|---|---|---|---|---|
| CDS | 13.7 | P<0.01 | 5.97 | 45 | 7 |
| Story | 9.2 | | 4.26 | 32 | 4 |

Both groups kept a downward trend in time to input the password and the trend in CDS group looked more salient. Moreover, the difference between two groups became more and more slight. For CDS group, an F-test (one tail) yields a result of F=4.11, p<0.01, showing a significant decrease in time to input the password over the ten logins. For the Story group, the result of F-test (one tail) indicated that there were no significant differences over the ten logins. It can be concluded that users will become adroit at CDS with frequent use.

## 4. COUCLUSIONS

A new shoulder-surfing resistant scheme CDS was proposed in this paper. It adopts a visual login technique that matches the capabilities and limitations of most handheld devices and provides a simple and intuitive way for users to authenticate on PDAs. The main contribution is that it overcomes a drawback of recall-based systems by erasing the drawing trace and introduces the drawing method to a variant of Story to resist shoulder-surfing. Preliminary usability testing of the CDS scheme showed that users were able to enter their passwords accurately. It took participants about 13.7 seconds to log in CDS at average and there was a gentle downward trend in time to input the password. The login time was not as good as Story and the degraded images in CDS were most likely to impinge on its efficiency.

Further work includes reducing the login time, finding a proper tolerance for the drawing trace length, and conducting a large scale user study. We also plan to investigate the entropy issue of pass-images and to study in more depth the effect of the association mnemonics on graphical passwords.

## 5. ACKNOWLEDGMENTS

The authors would like to thank the reviewers for their helpful and constructive comments. Project 60903198 supported by National Natural Science Foundation of China.